\documentclass[sn-mathphys-num]{sn-jnl}

\usepackage{graphicx}%
\usepackage{multirow}%
\usepackage{amsmath,amssymb,amsfonts}%
\usepackage{amsthm}%
\usepackage{mathrsfs}%
\usepackage[title]{appendix}%
\usepackage{xcolor}%
\usepackage{textcomp}%
\usepackage{manyfoot}%
\usepackage{booktabs}%
\usepackage{algorithm}%
\usepackage{algorithmicx}%
\usepackage{algpseudocode}%
\usepackage{listings}%
\usepackage[normalem]{ulem}

\newcommand{\ex}{\boldsymbol{\hat{e}}_{x}}
\newcommand{\ey}{\boldsymbol{\hat{e}}_y}
\newcommand{\ez}{\boldsymbol{\hat{e}}_z}
\newcommand{\es}{\boldsymbol{\hat{e}}_\text{s}}
\newcommand{\en}{\boldsymbol{\hat{e}}_\text{n}}

\newcommand{\xbb}{\boldsymbol{x}}

\newcommand{\beq}{\begin{equation}}
\newcommand{\eeq}{ \end{equation}}
\newcommand{\beqa}{\begin{eqnarray}}
\newcommand{\eeqa}{ \end{eqnarray}}

\theoremstyle{thmstyleone}%
\theoremstyle{thmstyletwo}%

\theoremstyle{thmstylethree}%

\begin{document}

\title{Effect of thermal fluctuations on the average shape of a graphene nanosheet suspended in a shear flow}


\author*[1]{\fnm{Simon} \sur{Gravelle}}\email{simon.gravelle@cnrs.fr}

\author[2]{\fnm{Catherine} \sur{Kamal}}\email{ck620@cam.ac.uk}
\equalcont{These authors contributed equally to this work.}

\author[3]{\fnm{Lorenzo} \sur{Botto}}\email{l.botto@tudelft.nl}
\equalcont{These authors contributed equally to this work.}

\affil*[1]{\orgname{Univ. Grenoble Alpes, CNRS, LIPhy}, \postcode{38000} \city{Grenoble}, \country{France}}

\affil[2]{\orgdiv{Cambridge Graphene Centre}, \orgname{University of Cambridge},  \city{Cambridge}, \postcode{ CB3 0FA}, \country{UK}}

\affil[3]{\orgdiv{Process and Energy Department, 3ME Faculty of Mechanical, Maritime and Materials Engineering}, \orgname{TU Delft}, \city{Delft},  \country{The Netherlands}}

\abstract{Graphene nanosheets display large hydrodynamic slip lengths in most solvents, and because of this, adopt a stable orientation in a shear flow instead of rotating when thermal fluctuations are negligible [{Kamal et al., Nature Comm., 11.1, 2020}]. In this paper, we combine molecular dynamics simulations and boundary integral simulations to demonstrate that the time-averaged `S' shape adopted by a flexible graphene nanosheet subject to moderate thermal fluctuation is comparable to the shape predicted when neglecting thermal fluctuations. The stable `S' shape adopted by the particle results primarily from the normal hydrodynamic traction, which is sensitive to the orientation of the particle with respect to the flow direction. Our results imply that thermally-induced shape fluctuations have a relatively minor effect on the time-averaged rheology of dilute suspensions of graphene nanosheets for relatively large but finite P\'{e}clet numbers.}



\maketitle

\section{Introduction}\label{sec1}

Graphene and other sheet-like nanomaterials have recently attracted the interest of the fluid dynamics community \cite{liu2017graphene, hamze2021graphene}. These are nanometrically thin, sheet-like particles which can deform under sufficiently large bending loads. In many applications, such as electronics \cite{avouris2012graphene}, energy  \cite{brownson2011overview}, and biomedical technologies \cite{chung2013biomedical}, these particles are processed as colloidal dispersions \cite{hernandez2008high, rao2015comparative,lalwani2013two}. The propensity for deformation under flow of these flexible particles and the rheological behavior of inks made with graphene (or other 2D materials) have therefore become of interest from both the applied and fundamental perspectives \cite{yu2022wrinkling, perrin2023hydrodynamic, salussolia2022simulation, yu2021coil, giudice2018filling, stafford2021real}. 

In previous studies, we have investigated \emph{rigid} nanosheets of pristine graphene in simple shear flow via molecular dynamics (MD) and continuum boundary integral (BI) simulations \cite{kamal2020hydrodynamic}. For rigid sheets, we have demonstrated that for slip lengths $\lambda \gtrapprox b$, where $b$ is the particle half-thickness, an elongated sheet suspended in a simple shear flow aligns at a constant small angle with the flow direction \cite{kamal2020hydrodynamic, gravelle2021violations}, rather than rotating as expected for elongated particles without slip \cite{jeffery1922motion}. For no-slip particles, stable alignment has been demonstrated for particles of a particularly complex shape, such as rigid ring-like particles with a non-circular cross-section of the curved solid \cite{singh_koch_stroock_2013,borker2024robust} or boomerang-shaped rigid rods \cite{roggeveen2022motion}. However, for slip particles, the condition for alignment only requires a plate like geometry with a small thickness and a slip length greater than the particle thickness \cite{kamal2020hydrodynamic, gravelle2021violations}. Since graphene is known to display a relatively large hydrodynamic slip length $\lambda$ in many solvents, \cite{maali2008measurement, tocci2014friction}, much larger than the thickness of the particle $2 b$, an alignment of a graphene particle in a shear flow is to be expected.  The importance of this observation stems from the fact that non-rotating slip disk-like particles, stably aligned almost parallel to the flow direction, yield a smaller effective viscosity than rotating particles with an average orientation in the flow direction. For dilutes suspensions of slip platelets at high P\'eclet numbers, the effective shear viscosity of the suspension can be even smaller than the suspending fluid viscosity, an anomaly among particulate suspensions \cite{kamal2023effect,kamal2024flow}. 

The criterion for alignment, however, depends on the average magnitude of the thermal or particle-induced fluctuations of the particle \cite{kamal2021effect, gravelle2021violations}. If the particle's fluctuation is above a certain threshold, the particle's dynamic will change from stable, where the particle fluctuates about its stable orientation, to unstable, where the particle tumbles.  So far, studies to predict the threshold fluctuation for tumbling, for the case of Navier-slip particles, have focused on rigid particles \cite{kamal2021effect}. However, for a thin particle in suspension, the assumption of rigidity is only valid for relatively small aspect ratios and applied shear stresses \cite{kamal2021alignment}. In the case of a monoatomic graphene particle, whose thickness is subnanometer, flexibility and Brownian motion can induce undulations in the shape of the particle, which in turn can affect the particle's orientational distribution. It has been demonstrated using continuum calculations that in the absence of thermal fluctuations, a flexible sheet with hydrodynamic slip assumes a `S' shape with an orientation angle that is similar to the one obtained by assuming the sheet to be rigid \cite{kamal2021alignment}. However, the effect of flexibility on the attainment of this stable orientation is not completely understood, in particular in the presence of Brownian fluctuations (and possibly other atomistic effects related to the discrete nature of the fluid molecules). An increase in shape deformation may change the stability region of the particle when subject to thermal or flow-induced shape fluctuations. 

In this paper, we combine MD and BI simulations to quantify the average shape of a flexible graphene particle, and assess whether Brownian fluctuations in the particle shape are sufficient to induce a reorientation of the particle. The particle is a graphene nanoplatelet with bending rigidity either equal to the one of pristine graphene or an artificially reduced bending rigidity, where the reduction factor is denoted $\alpha$. This modulation of the bending rigidity allows investigating deformation effects while still simulating a small particle, a constraint imposed by the use of MD.  The liquid is water. The time-averaged shapes extracted from MD for different particle bending rigidity are compared to a model based on the Bernoulli beam equation with hydrodynamic stresses extracted from BI. Our study extends the knowledge accumulated on the effect of Brownian motion of flexible fibers \cite{liu2018morphological, du2019dynamics}, i.e., the most extensively studied anisotropic particles, to the case of 2D materials such as graphene. In studies on flexible fibers, the comparison is usually made between continuum simulations and experiments. Here, the comparison is between MD and continuum simulations. We believe that comparison against MD is necessary because the thickness of graphene is comparable to the size of the suspending liquid molecules, so the validity of the continuum assumption needs to be tested.

The MD simulations are carried out for statistically two-dimensional cases in which the sheets have an infinite extent in the vorticity direction. This approach is taken for computational convenience - it allows us to accurately estimate the viscous traction by averaging over the homogeneous direction and to keep the computational cost manageable by using domains with limited extent in the spanwise direction - and for amenability to mathematical analysis. We note that previous MD simulations \cite{gravelle2021violations} of a fully three-dimensional disk-like nanographene sheet, for values of the P\'eclet number comparable to those used in the present work, show that the particle's axis of rotation exhibited a random motion in the orientation space of the polar and azimuthal angles (with the polar axis along the vorticity direction). In these simulations, in which the disk exhibited three-dimensional shape deformations, the probability distribution of the polar angle was found to be symmetric, resulting in an average polar angle equal to zero. This result corresponds to a time-averaged planar motion in the flow-gradient plane.  

\subsection{Simulation details} 

\begin{figure}
\centering
\includegraphics[width=\textwidth]{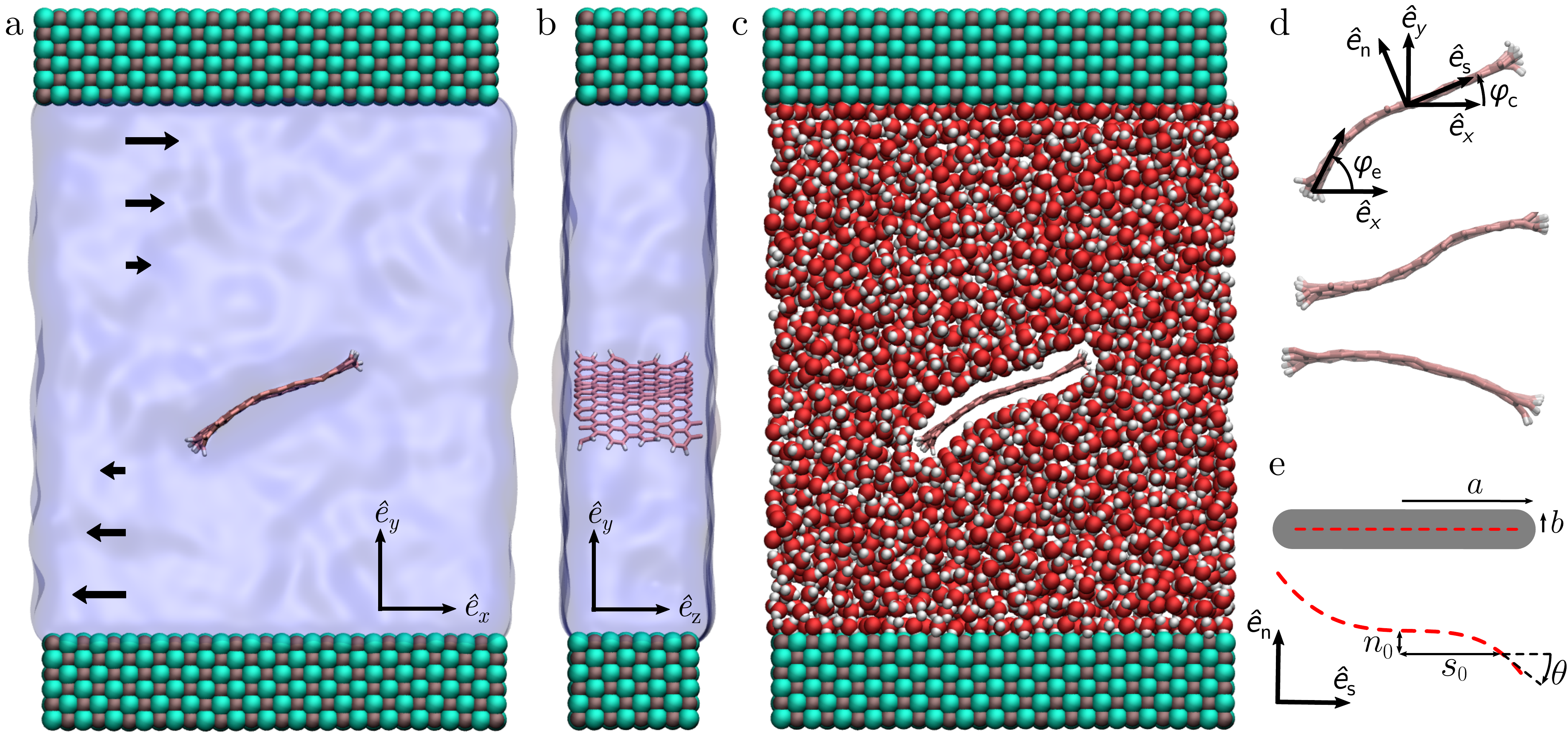}
\caption{a-c) Molecular dynamics system with a single graphene particle immersed in water \cite{humphrey1996vmd}. The graphene particle has its orientation confined to the flow-gradient plane ($\ex$, $\ey$). The two layers at the top and bottom are the shearing walls, and the undisturbed linear shear flow is $u_{x} = \dot \gamma y$. Water is represented as a transparent continuum field (a-b) and as red and white spheres (c). The size of the graphene particle along the $\ez$ direction, $w = 1.7\,$nm, corresponds to the dimension of the computational domain (b).
d) View of the particle at different times. The top-most snapshot shows the coordinate system with the two angles $\varphi_\text{c}$ and $\varphi_\text{e}$. The frame of the laboratory is ($\ex$, $\ey$), and the particle frame of reference is ($\es$, $\en$) rotates with the platelet. The tangent to the platelet at the center of the platelet, $\es$, is inclined by an angle $\varphi_\text{c}$ with respect to $\ex$. The tangent to the platelet at the edge of the platelet is inclined by an angle $\varphi_\text{e}$ with respect to $\ex$.
e) Schematic of the undeformed (top) and deformed (bottom) flexible particle of half-length $a$ and half-thickness $b$ used for the continuum calculations. The red dashed line is the particle centerline, $n_0$ and $s_0$ are the coordinates of the centerline within the particle frame of reference ($\es$, $\en$), and $\theta$ is the angle between the tangent of the centerline and $\es$.}
\label{fig:Figure1}
\end{figure}

\noindent \textit{Molecular Dynamics simulations.} MD simulations of a freely suspended graphene particle in a shear flow were performed using LAMMPS \cite{thompson2022lammps}. The particle was made of a single graphene layer of approximate length $2 a=3.2$\,nm and width $w=1.7$\,nm, with hydrogen atoms terminating the carbon atoms at the edges. The spanwise dimension of the computational domain in the $\ez$ direction was equal to the width of the particle, $w$, and periodic boundary conditions were used in the three dimensions of the space. In this quasi-2D configuration, the particle was free to move within the ($\ex$, $\ey$) plane and to rotate around the $\ez$ axis. The fluid was made of a number $N=2500$ water molecules, enclosed in the $\ey$ direction by two moving walls made of iron(II) oxide (FeO) separated by an average distance of 6\,nm (Figure~\ref{fig:Figure1}\,a-c). FeO was chosen for its high friction coefficient, which corresponds to a slip length $\lambda \approx 0$\,nm in water. A linear shear flow $u_x = \dot \gamma y$ with $\dot \gamma = 10^{10}$\,s$^{-1}$ was generated by translating the two walls along the $\ex$ direction. The walls were also used to impose the atmospheric pressure $p=1$\,atm to the fluid, thanks to an additional forcing applied in the $\ey$ direction. The TIP4P/2005 model was used for water \cite{abascal2005general}. For the unmodified graphene particle, the all-atom GROMOS force field was used \cite{schmid2011definition}. Particles with reduced bending rigidity were also created by multiplying the dihedral constraints by a factor $1/\alpha$, with $\alpha = 5$, $10$, and $100$. The crossed parameters of the Lennard-Jones interaction were calculated using the Lorentz-Berthelot mixing rule. The moving walls and the water were both maintained at a temperature $T=300$\,K thanks to a Berendsen thermostat applied only to the degree of freedom normal to the direction of the flow, and with a time constant of 100\,fs. The input and topology files are available in the GitHub repository; see the data availability statement.

\textit{Particle orientation from MD.} Using the previously described system, production runs of duration 12\,ns were performed (for each particle, i.e. for each value of $\alpha$). During the production runs, the particle configuration was saved every picosecond. For each recorded frame, the instantaneous orientation of the tangent to the particle surface at its center with respect to the laboratory frame ($\ex$, $\ey$), $\varphi_\text{c} (t)$, was extracted. Then, the instantaneous shape of the particle was analyzed within the particle frame of reference ($\es$, $\en$) rotated by $\varphi_\text{c} (t)$ relative to the laboratory frame (Figure~\ref{fig:Figure1}).

\textit{Bending rigidity from MD.} The bending rigidity values were measured using loading MD simulations performed in water \cite{wang2010simulations}. The particles were initially maintained flat in the ($\ex$, $\ez$) plane by preventing the motion of the hydrogen atoms at the edge in the $\ey$ direction. Then, a constant force of maximum magnitude $3$\,kcal/mol/\AA{} was applied along $\ey$ to the atoms located at the center of the platelet, to induce a curved shape in the platelet with curvature $\kappa$. Both curvature $\kappa$ and potential energy $E_\text{p}$ of the platelet were measured, and the bending rigidity was calculated as $B = 2 E_\text{p} / (\kappa^2 {\cal A})$ \cite{lambin2014elastic}, where ${\cal A} = 2 (a - b) w$ is the surface area of the platelet, excluded in the edge of size $b$. Our results for the bending rigidity of pure graphene ($\alpha=1$) is $B = 1.6$\,eV, in good agreement with previous simulations \cite{lu2009elastic, wang2010simulations, wei2013bending} and experiments \cite{nicklow1972lattice}. For flexible platelets with increasing factor $\alpha$, our results are $B = 0.43$\,eV ($\alpha = 5$), $B = 0.28$\,eV ($\alpha = 10$), and $B = 0.14$\,eV ($\alpha = 100$). Therefore, the ratio between the bending energy and the thermal energy, $k_\text{B} T = 0.0258$\,eV, where $k_\text{B}$ is the Boltzmann constant and $T = 300$\,K the temperature, ranges from $0.016$ ($\alpha = 1$) to $0.185$ ($\alpha = 100$).

\vspace{0.5cm} \noindent \textit{Boundary Integral Simulations.} We compare MD simulations to continuum BI simulations of the two-dimensional incompressible Stokes equations. The continuum simulations are non-Brownian and consider an isolated two-dimensional rigid plate-like particle freely suspended in the linear shear flow $u_x=\dot{\gamma}y$. Such simulations have been shown to provide a good estimate of the hydrodynamic stress distribution calculated in MD provided that (i) a particular reference surface between the carbon and water molecules is used and (ii) a uniform hydrodynamic slip surface is used at the boundary \cite{kamal2020hydrodynamic, luo2008effect, kamal2021effect}. For single-layer graphene, a reasonable choice for the reference surface is the surface of a rectangle with rounded ends \cite{kamal2020hydrodynamic}. The cross-section of the reference surface consists of a rectangular central part of half-length equal to $a-b$ and thickness $b$ and two semi-circular edges of radius $b$, attached to each end of the rectangle (Figure~\ref{fig:Figure1}\,e). In our simulations, we used a maximum half-length $a=1.6$~nm and half-thickness $b=0.25$~nm, in accordance with the particle used in the MD simulations. For the slip surface, a Navier slip boundary condition was assumed with a Navier slip length $\lambda.$ The slip length was taken as $\lambda=10$\,nm, in agreement with experimental and atomic measurements of hydrodynamic slip of graphene in water \cite{maali2008measurement, tocci2014friction, kamal2020hydrodynamic, herrero2020fast}. The numerical method for obtaining the hydrodynamic traction is described in detail and verified in Refs.~\cite{kamal2020hydrodynamic, kamal2021effect}. Briefly, the hydrodynamic traction at the particle's stable orientation is calculated for an undeformed particle as follows. First, the stable orientation angle is determined by evaluating the torque $T(\varphi)$ acting on the particle when it is held fixed with its planar surface parallel ($\varphi = 0$) and perpendicular to the flow direction ($\varphi = \pi/2$). The torque $T(\varphi)$ is obtained by inverting the two-dimensional boundary integral equation numerically to obtain the hydrodynamic traction $\boldsymbol{f}$ of a two-dimensional particle oriented at the angle $\varphi$. The hydrodynamic traction is then used to compute $T(\varphi)=\int_{S} \boldsymbol{f}\times{\xbb} \mathrm{d}\ell$ where $S$ is the boundary of the particle, $\mathrm{d}\ell$ is a line element on this boundary and $\xbb$ corresponds to a point on $S$. The stable orientation angle $\left<\varphi_c\right>$, which for this case is independent of time for $\text{Pe} \to \infty$, corresponds to the value of $\varphi$ for which $\cos^{2}(\varphi) T(\varphi=0)+\sin^{2}(\varphi)T(\varphi=\pi/2)=0$ \cite{kamal2020hydrodynamic}. The hydrodynamic traction is then computed from the boundary integral equation for a particle oriented at $\left< \varphi_c \right>$. The hydrodynamic traction is used to compute the distributed hydrodynamic load given in the beam equation [See Eq.\,\eqref{eq:force_eqt} below].

\section{Results}

\begin{figure}
\centering
\includegraphics[width=0.8\linewidth]{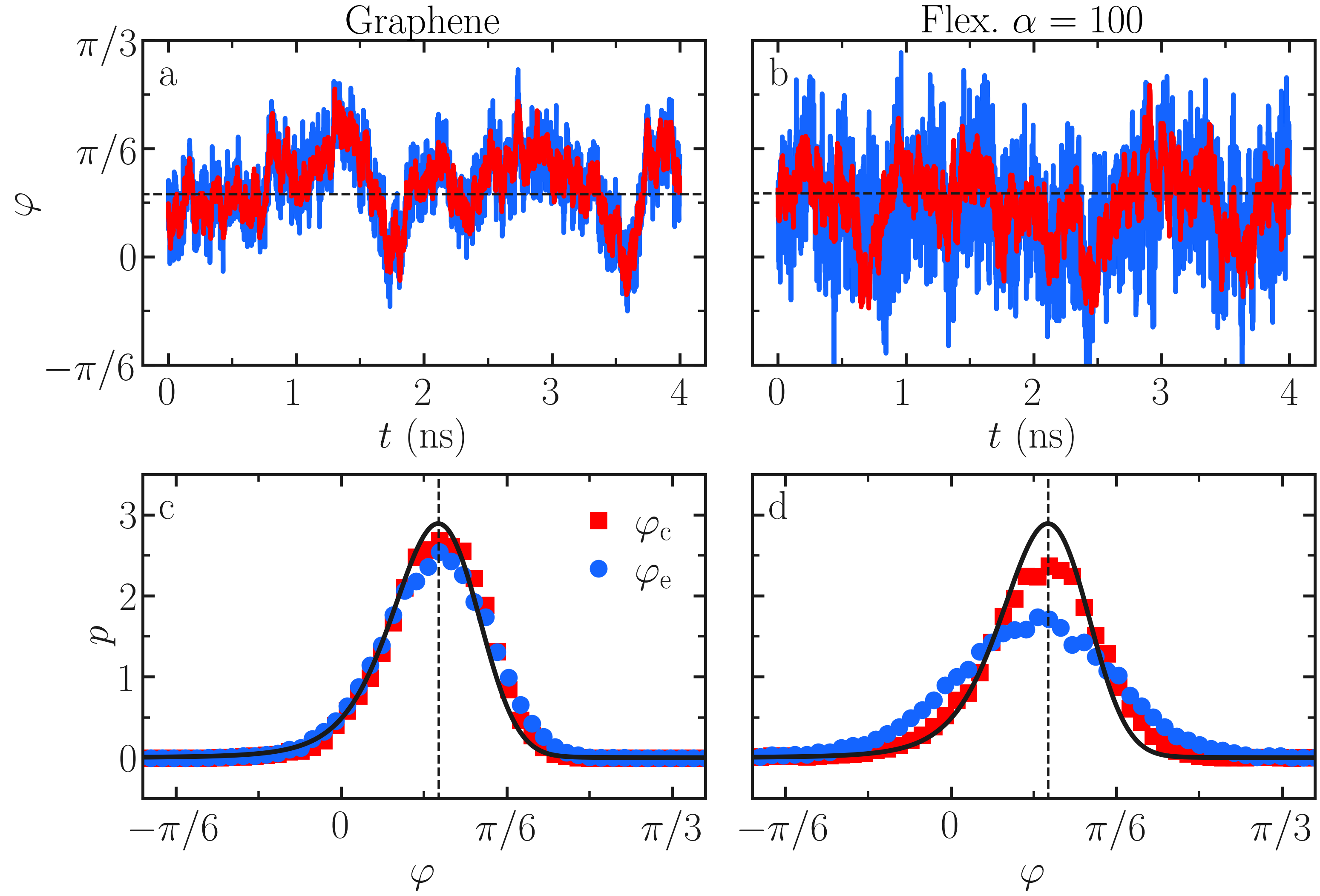}
\caption{
a-b) Angle at the center $\varphi_\text{c}$ (red) and at the edge $\varphi_\text{e}$ (blue) of the particle as a function of time $t$, for a graphene particle (a) and a flexible particle with $\alpha = 100$ (b). The dashed horizontal line is the most probable angle $\bar{\varphi} \approx 0.29$. c-d) Orientational distribution function corresponding to the mid-point angle (red squares) or edge angle (blue disks), for a graphene particle (c) and a flexible particle with $\alpha = 100$ (d). The vertical dashed line is the most probable angle $\varphi \approx 0.29$. {\color{blue}The blue and red symbols are results from MD, and }the full black line corresponds to the solution of Eqs.\,(\ref{FP_eq1D}-\ref{ke}) for a rigid particle with effective aspect ratio $k_\text{e}=0.4 i$ (i.e., a particle with large hydrodynamic slip) and rotational diffusion coefficient $D_\text{r}=1.2 \cdot 10^8$\,s$^{-1}$ (i.e., $\text{Pe} = 90$).} 
\label{fig:Figure2}
\end{figure}

The first aspect we address is how the flow affects the orientational statistics of the particle. As shown by the instantaneous configurations in Figure~\ref{fig:Figure1}, the particle attains an average orientation with respect to the laboratory axes, which are fixed with respect to the flow direction. Due to the sheet deformation, each element is characterized by its own rotation angle. To quantify the orientational statistics, we focus on two characteristic angles: the angle $\varphi_\text{c}$ between the tangent, $\es$, at the mid-point of the sheet and the $\ex$ axes, and the angle $\varphi_\text{e}$ between the tangent at one of the edges (the left one in Figure~\ref{fig:Figure1}\,d) and the $\ex$ axes. The time series of $\varphi_\text{c}$ and $\varphi_\text{e}$ are shown in Figs.\,\ref{fig:Figure2}\,a\,b. The corresponding probability distribution functions are shown in Figs.\,\ref{fig:Figure2}\,c\,d. Panels a and c in Figure~\ref{fig:Figure2} show data for the pristine graphene sheet, while panels b and d are for the graphene sheet with artificially reduced bending rigidity, $\alpha = 100$.

The time series shows that $\varphi_\text{c}$ and $\varphi_\text{e}$ fluctuate in time around a time-averaged angle $\left<\varphi\right>=0.25$. The most probable angle in time is $\bar{\varphi} = 0.29$. As the bending rigidity decreases, the amplitude of the fluctuations increases, and the correlation between the fluctuations in $\varphi_\text{c}$ and $\varphi_\text{e}$ decreases. The orientational probability distribution functions,  which are symmetric about the average, are essentially zero outside the range $[-\pi/6, \pi/3]$. If the sheet performed a tumbling motion, the probability would be nonzero for any value of $\varphi_\text{c}$, so we infer that our sheets do not tumble. This behavior is confirmed by visual monitoring of the sheets in the MD simulations. Our results also reveal that $p(\varphi_\text{e})$ is slightly broader than $p(\varphi_\text{c})$, which is particularly visible for the most flexible particles considered here (Figure\,\ref{fig:Figure2}\,c-d).

A suppression of tumbling motion is expected from the theory developed for rigid sheets \cite{kamal2020hydrodynamic, gravelle2020liquid, gravelle2021violations}. According to this theory, at sufficiently high rotational P\'eclet number, $\textrm{Pe}$, a rigid sheet is expected to remain aligned at a small constant angle $\left<\varphi\right> = \varphi^*$ with respect to the flow direction when the hydrodynamic slip length $\lambda$ is approximately larger than the half-thickness $b$ of the particle. For graphene in water, the slip length is relatively large, typically $\lambda \ge 10$\,nm \cite{maali2008measurement, tocci2014friction, herrero2020fast}, while $b \approx 0.3-0.4$\,nm \cite{kamal2020hydrodynamic} for single-layer graphene. So, for sufficiently high $\text{Pe}$, our sheets are predicted to maintain a stable orientation, provided that the sheet deformation is sufficiently small for the rigid sheet theory to hold.

For sheets whose shape does not deviate significantly from a straight sheet, a model for the orientational distribution function $p$ at finite $\textrm{Pe}$ can be developed using the theory for rigid sheets. The basic ingredients of such a model are a diffusive flux, $ D_\text{r} \partial p/\partial \varphi$, which tends to bring $\varphi$ away from the equilibrium angle $\varphi^*$, and a convective flux of magnitude $\Omega (\varphi) \dot{\gamma} p$ which tends to bring $\varphi$ towards $\varphi^*$ \cite{gravelle2021violations}. The corresponding Fokker-Planck equation reads 
\begin{equation}
\dfrac{\partial p}{\partial t} = 
\dfrac{\partial}{\partial \varphi} \left( D_\text{r} \dfrac{\partial p}{\partial \varphi} 
- \Omega (\varphi) \dot{\gamma} p \right),
\label{FP_eq1D}
\end{equation}
where $\Omega (\varphi) \dot{\gamma}$ is the hydrodynamic angular velocity, $\dot{\gamma}$ is the shear rate, and $D_\text{r}$ the rotational diffusion coefficient of the particle \cite{leahy2015effect}. We seek steady solutions of this equation. For a rigid anisotropic particle, a classical solution due to Bretherton \cite{bretherton1962motion, kim2013microhydrodynamics} provides an exact expression for $\Omega (\varphi)$ as a function of a scalar parameter $k_\text{e}$:
\begin{equation}
\Omega (\varphi) = - \left(k_\text{e}^2\cos{\varphi}^2+\sin{\varphi}^2\right) / (1 + k_\text{e}^2).
\label{jeffery_vel}
\end{equation}
In most studies of anisotropic particles, $k_\text{e}$ is regarded as a real number. However, if  $k_\text{e}$ is real, no steady solution can be found for $\textrm{Pe} \to \infty$ (as can be seen by setting $D_r = 0$ in Eq.\,\eqref{FP_eq1D} and noting that $\Omega (\varphi)<0$ for any $\varphi$ if $k_\text{e}$ is real). A steady solution for $\textrm{Pe} \to \infty$ and fluctuations around the steady solution for finite $\textrm{Pe}$ require $k_\text{e}$ to be complex imaginary so that $\Omega (\varphi)$ admits a zero. An imaginary solution can occur if the slip length is sufficiently large, the case considered in the current paper. 

The parameter $k_\text{e}$ is defined as  \cite{cox1970motion}
\begin{equation}
k_\text{e}=\sqrt{\frac{T(\varphi = 0)}{T(\varphi = \pi/2)}},
\label{ke}
\end{equation}
where $T(\varphi = 0)$ and $T(\varphi = \pi/2)$ are the hydrodynamic torques on a particle held fixed parallel and perpendicular to the flow, respectively. To compare the theory [Eqs.\,(\ref{FP_eq1D}-\ref{ke})] with the MD data, $k_\text{e}$ was calculated from the MD simulations by measuring the torque applied by the fluid on a rigid graphene nanosheet oriented at either $\varphi=0$ or $\pi/2$. With a shear rate $\dot \gamma = 10^{10}$\,s$^{-1}$, time-averaged torques of $-(6.9 \pm 0.9)$\,kcal/mol and $(36.6 \pm 0.9)$\,kcal/mol were measured for respectively $\varphi=0$ and $\varphi=\pi/2$, and a value $k_\text{e} = (0.4 \pm 0.1 ) i$ was calculated using Eq.\,\eqref{ke}.  

We solved Equation~\eqref{FP_eq1D} for $k_\text{e}=0.4 i$ and $\text{Pe}=90$ using the spectral method described in Ref.~\cite{kamal2021effect}. An excellent agreement was found between the MD simulations and the theory [Eqs.\,(\ref{FP_eq1D}-\ref{ke})] (Figure\,\ref{fig:Figure2}\,c-d). The main difference between the model and the MD results is that the variance of the angular distribution from MD is slightly larger than the variance predicted by Eqs.\,(\ref{FP_eq1D}-\ref{ke}), particularly for the most flexible particle considered here ($\alpha = 100$). The value of 90 used for the P\'eclet number when solving Eqs.\,(\ref{FP_eq1D}-\ref{ke}) corresponds to a rotational diffusion coefficient $D_\text{r} = \dot \gamma / \text{Pe} = 1.1 \cdot 10^8$\,s$^{-1}$ with $\dot \gamma = 10^{10}$\,s$^{-1}$. This value is in excellent agreement with the estimate $D_\text{r} \approx 3 k_\text{B} T / 32 \eta a^3 \approx  1.1 \cdot 10^8$\,s$^{-1}$ for an infinitely thin rigid disk of radius $a=1.6$\,nm, where $k_\text{B}$ is the Boltzmann constant and $T$ the temperature \cite{sherwood2012resistance}. Therefore, the comparison between the model prediction and the data is obtained with no free parameters. The only parameter, $k_\text{e}$, was obtained from MD. In principle, $k_\text{e}$ could also be calculated from our boundary integral simulations, assuming the particle is rigid \cite{kamal2020hydrodynamic}. By doing so, we find $k_\text{e}=0.4i$, which agrees with the MD value.

\begin{figure}
\centering
\includegraphics[width=0.8\linewidth]{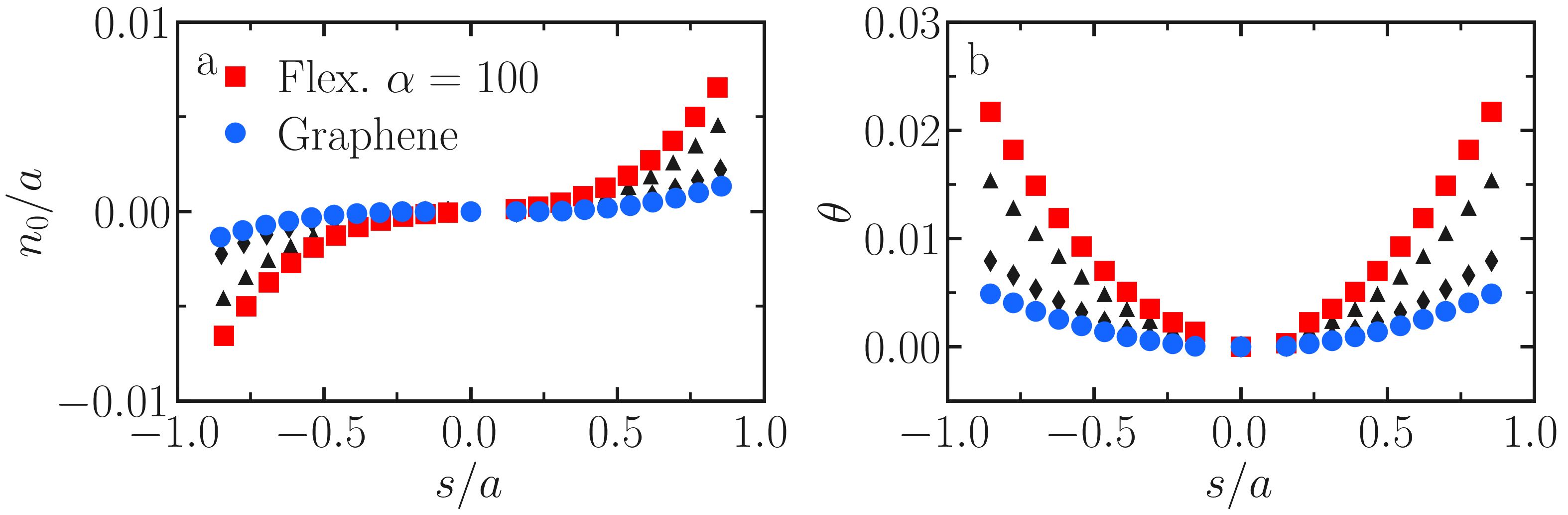}
\caption{(a) Time-averaged profiles in the particle reference frame ($\es$, $\en$). The data corresponds to an unmodified graphene particle (blue disks), and a modified particle with reduced bending rigidity,  with $\alpha=5$ (gray diamonds), $\alpha=10$ (gray triangles), and $\alpha=100$ (red squares).
(b) Time-averaged rotation angle $\theta$ corresponding to the data in panel a.}
\label{fig:Figure3}
\end{figure}

We now examine the time-averaged shape of the sheet, obtained by averaging the positions of the solid atoms from the MD results within the particle frame of reference ($\es$, $\en$). The profiles of the averaged longitudinal centreline of the sheet $\boldsymbol{x}_0=(s_0 ,\;n_0)$ show a characteristic `S' shape (Figure~\ref{fig:Figure3}\,a), with the largest average deformations obtained for the most flexible sheet considered ($\alpha = 100$). We then extract the rotation angle, $\theta$, between the tangent to the centerline and the $\es$ axis (see Figure~\ref{fig:Figure1}\,e).
The rotation angle $\theta$, is maximum at the edges, and the more flexible the sheet, the larger the maximum value of $\theta$ (Figure~\ref{fig:Figure3}\,b).

The time-averaged profile at $\text{Pe} \simeq 90$ displays an S-shape. But so does the steady profile at $\text{Pe} =  \infty$ \cite{kamal2021alignment}. It is therefore natural to attempt to model the time-averaged shape of the sheet using the hydrodynamic load obtained from steady BI simulations at $\text{Pe} \to  \infty$. To do so, we first parameterize the center line by its arc-length $\{s:\;-a\leq s \leq a \}$. In terms of $s$, $n_0$ and $s_0$ obey the kinematic relations
\begin{equation}
\label{eq:parametrization}
\dfrac{\partial n_0}{\partial s} = - \sin (\theta),
\hspace{0.5cm}
\dfrac{\partial s_0}{\partial s} = \cos (\theta),
\hspace{0.5cm}
\dfrac{\partial \theta}{\partial s} = \kappa,
\end{equation}
where $\kappa$ is the curvature and $\theta =0$  at the sheet's midpoint $s=0$. Treating the flexible object as a slender linear elastic object when observed from the flow-gradient plane, the governing equation for $\theta$ is \cite[]{audoly2010elasticity,kamal2021alignment} 
\begin{equation}
\hat{B} \dfrac{\partial^3 \theta}{\partial s^3}  = g_{\perp} + \dfrac{\partial q}{\partial s},
\label{eq:force_eqt}
\end{equation}
where $g_{\perp} (s)$ is the average surface traction directed in the normal direction to the center line, and $q (s)$ is the torque density applied by the fluid to the particle, respectively. The parameter
\beq
\hat{B}= \frac{B}{ \dot{\gamma} \eta L^3}
\label{eq:bending_rigidty}
\eeq
is the non-dimensional bending rigidity for a plate \cite{kamal2021alignment}, also referred to as the elasto-viscous number \cite[]{silmore2021buckling,perrin2023hydrodynamic}. Treating graphene as a homogeneous solid, $B\approx B_0 b^3 $ \cite{poot2008nanomechanical,landau1995theory}. In our simulations, we set $B_0\approx 2.3 \times 10^{11}~\text{N~m}^{-2}$, which is inline with reported values of bending rigidity for graphene \cite{lindahl2012determination}.  Equation~\eqref{eq:force_eqt} is similar to the large-bending deformation model used for a 1D fiber \cite{du2019dynamics}, and is equivalent to the Bernoulli beam equation when the deformation is small \cite[]{audoly2010elasticity,kamal2021alignment}. The difference with equivalent models for fibers is with the computation of the hydrodynamic forcing, $g_{\perp} + \partial q/\partial s$. In our case, we show below that the forcing can be computed from the BI simulations described in the `simulations details' section, and not from a slender body approximation as for the case of fibres \cite{du2019dynamics}.

The difficulty in predicting the shape of the centerline of the sheet lies in the prescription of the loads $g_\perp$ and $q$, as well as, the edge loads. To distinguish the edge load from the load on the rest of the structure, we split the centerline into two regions: a slender region of length $2L$ for $ -L\leq s \leq L$ and `edge' regions for $L< |s| \leq a$ \cite[]{pozrikidis2010shear,kamal2021alignment}. Defining the half thickness $b$, then $L=a-b$. The slender region is assumed to satisfy the Bernoulli beam equation, with external force and torques applied from the edge region. Equation~\eqref{eq:force_eqt} is supplemented with the following boundary conditions:
\begin{eqnarray}
\nonumber
\hat{B} \left. \dfrac{\partial^2 \theta}{\partial s^2} \right|_{s=\pm L}&=&-q(\pm L) + F_{\text{e}, \perp}(\pm L),\\
\label{eq:bc}
\hat{B} \left. \dfrac{\partial \theta}{\partial s} \right|_{s=\pm L}&=&T_{\text{e}}(\pm L),\\
\nonumber
\left. \theta \right|_{s = 0} &=& 0,
\end{eqnarray}
where $F_{\text{e},\perp}(\pm L)$ and $T_{\text{e}}(\pm L)$ correspond to the applied normal force and torque from each edge, respectively.

To solve Eqs.\,(\ref{eq:force_eqt}-\ref{eq:bc}), values for the distributed and edge loads  $g_{\perp}$, $q$, $F_{\text{e},\perp}$ and $T_{\text{e}}$ are needed. We have calculated these values from our BI simulations for a straight (undeformed) geometry, solving for the hydrodynamic traction around the rigid particle oriented at its stable orientation $\tan{\varphi}=|k_\text{e}|$ \cite{kamal2020hydrodynamic}.  {The details of these simulations are given in the `Simulation Details' section.}  

The calculation of $g_{\perp}$ and $q$ proceeds as follows. The surface was parametrized as 
$$ \{ S = \left[ s_0(s) \pm h(s) \sin\theta(s), n_0(s) \pm h(s) \cos \theta(s) \right] : \; -a \leq s \leq a \} ,$$
where $h(s)$ is the half-thickness of the particles.  The surface traction $\boldsymbol{f}(s,\pm h)$, is averaged over the center line by considering 
\begin{equation}
\boldsymbol{g}=\boldsymbol{f}({s},{h}) + \boldsymbol{f}({s},-{h}), \quad \Delta \boldsymbol{g}({s},{h})  =  \boldsymbol{f}({s},{h}) - \boldsymbol{f}({s},-{h}).
\end{equation}
Defining the normal and tangent vector to the center line by $\boldsymbol{n}(s)$ and $\boldsymbol{t}(s)$, respectively,  the normal traction component $g_{\perp} (s)=\boldsymbol{g} (s)\cdot\boldsymbol{n}(s)$, and the torque density is
\beq
{q}({s}) =  {h} ({s}) \left[ \Delta g_{\perp}({s})\sin{\theta({s})}-\Delta g_{\parallel}({s})\cos{\theta({s})} \right],
\label{torque_density}
\eeq
where $\Delta{g}_{\perp}(s)=\Delta{\boldsymbol{g}} (s)\cdot\boldsymbol{n}(s)$ and $\Delta{g}_{\parallel} (s)=\Delta{\boldsymbol{g} (s)}\cdot\boldsymbol{t}(s)$.
The edge forces and torques $F_{\text{e},\perp}$ and $T_{\text{e}}$ were calculated by integration of the stress over the edge surfaces $S_E(-L)$ and $S_E(L)$:
\beq
T_{e}(\pm L) = \ez \cdot \int_{S_{E}(\pm L)}\left(\boldsymbol{x}-\boldsymbol{x}_{e} \right)\times\boldsymbol{f} \mathrm{d}S, 
\quad F_{e,\perp}(\pm L)=\int_{S_{E}(\pm L)}{f_{\perp}} \mathrm{d}S,
\label{ends}
\eeq
where $\boldsymbol{x}_{e}$ is the end point of the slender region (i.e., $s=\pm L$). 

\begin{figure}
\centering
\includegraphics[width=\linewidth]{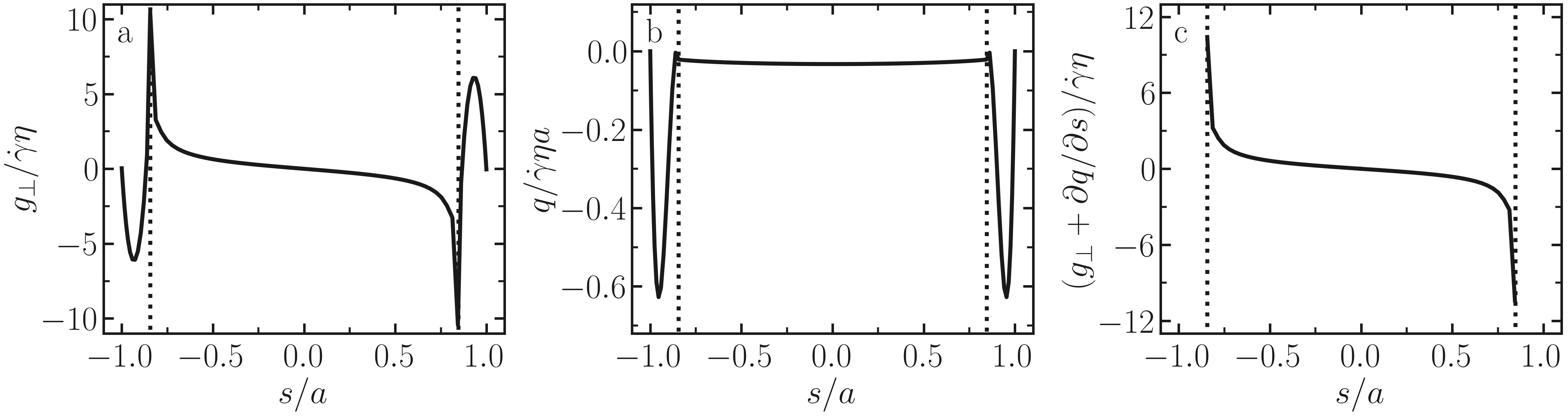}
\caption{Hydrodynamic load applied by the fluid on the particle, as computed via BI for a rigid particle. Normal component of the traction $g_{\perp}$ normalized by $\dot \gamma \eta$ (a), torque density $q$ normalized by $\dot \gamma \eta a$ (b), and hydrodynamic load $g_{\perp} + \partial q /\partial s$ normalized by $\dot \gamma \eta$ (c) as a function of the coordinate $s$. The particle is oriented at  $\varphi = 0.3$. The vertical dotted lines represent the boundaries between the slender region and the edge region.}
\label{fig:Figure4}
\end{figure}

The computed values of $g_{\perp}$, $q$, and $g_{\perp}+ \partial q /\partial s$ are shown in Figure~\ref{fig:Figure4}. The hydrodynamic load is primarily dominated by $g_{\perp}$. This function varies roughly linearly with $s$ away from the edges and peaks near the edges. Since slip affects the hydrodynamic traction in the tangential direction but not in the normal direction, the normal load distribution we report in Figure~\ref{fig:Figure4} is qualitatively similar to that for no-slip particles oriented in the flow direction \cite{kamal2020hydrodynamic}. This explains why a similar distribution of $g_{\perp}$ with $s$ is observed for no-slip elongated particles that are oriented with their major axes parallel to the flow direction  \cite[]{singh2014rotational,kamal2020hydrodynamic,zuk2021universal}.

\begin{figure}
\centering
\includegraphics[width=0.8\linewidth]{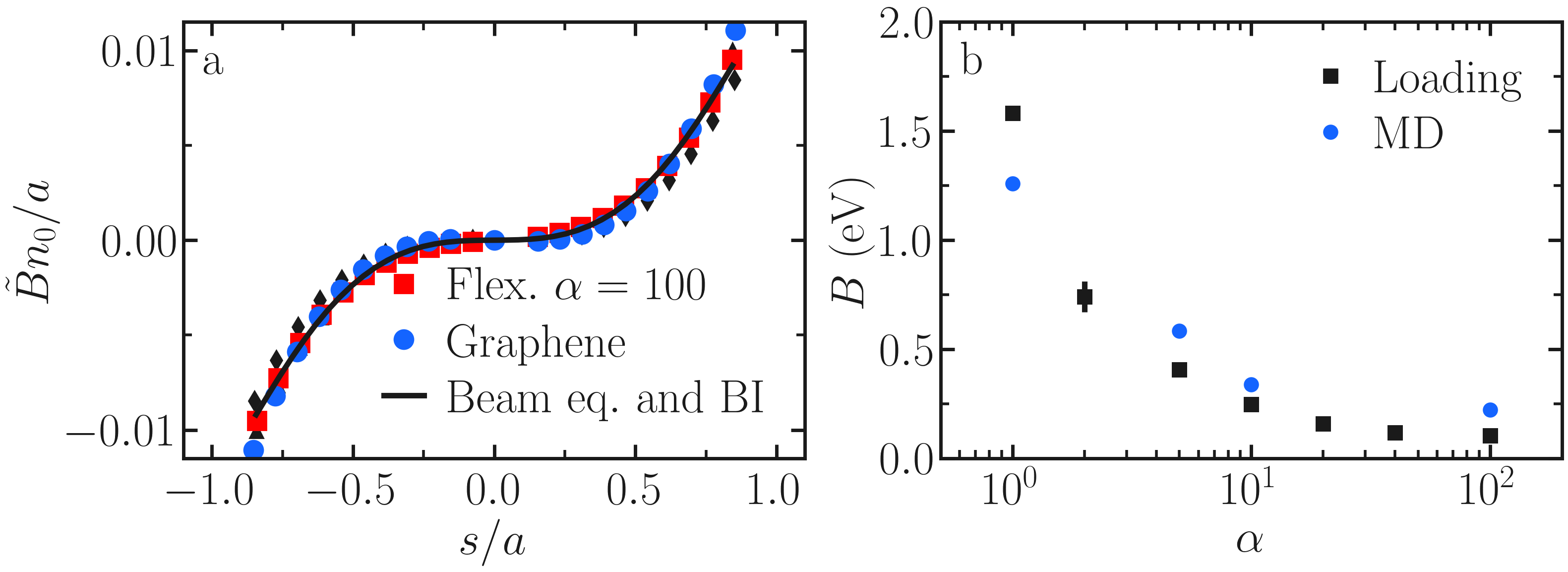}
\caption{a) Comparison of rescaled deflection $\tilde{B}n_0 / a$ from the continuum calculations (Beam eq. + BI, full line) and from the MD simulations (symbols). The MD data corresponds to an unmodified graphene particle (blue disks), and a modified particle with decreased bending rigidity, with $\alpha=5$ (gray diamonds), $\alpha=10$ (gray triangles), and $\alpha=100$ (red squares). The continuum calculations correspond to the solution of the beam equation [Eq.\,\eqref{eq:force_eqt}] with the loads ($g_\perp$, $q$, $F_{\text{e}, \perp}$, and $T_\text{e}$) calculated using BI, see the text for details.
b) Bending rigidity $B$ of the particles as a function of the factor $\alpha$, where $\alpha = 1$ corresponds to pristine graphene, obtained from direct loading measurements (black squares; see `Simulation details') and by matching MD profiles in panel (a) to the continuum model (blue disks).}
\label{fig:Figure5}
\end{figure}

The computed hydrodynamic load is used in Eqs.\,\eqref{eq:force_eqt} and \eqref{eq:bc} to find $\theta$, and the corresponding deformed shape of the sheet is computed from Eq.\,\eqref{eq:parametrization}. A finite difference method is used for discretization. The normal deflection of the sheet is shown as a solid curve in Figure~\ref{fig:Figure5}\,a. The deflection has been rescaled by a factor $\hat{B}$ to show that the average deflections of sheets of different bending rigidity are identical up to a multiplicative prefactor, and that the shapes given by the continuum and MD simulations are perfectly overlapping. Furthermore, the values of  $\hat{B}$ that give collapse onto a single curve are in agreement with direct measurements of the bending rigidity using the loading simulations described in the method section (Figure~\ref{fig:Figure5}\,b). Because $\hat{B}$ is numerically close to the value obtained from independent MD measurements of the bending rigidity, the agreement between the continuum solution of the beam equations and the MD data for the sheet deflection can be considered as achieved without resorting to free parameters.

\section{Conclusions}
\label{sec12}
We have presented the first direct comparison between Molecular Dynamics and continuum simulations of the time-averaged shape of a flexible graphene nanosheet in shear flow. The graphene sheet has a slip length larger than its thickness, and therefore does not perform a full tumbling motion but fluctuates about a preferred orientation angle $\left<\varphi_c\right>$. In this case, $\left<\varphi_c\right> = 0.25$, meaning the particle is nearly aligned with the flow direction. The main result is that the time-average of the fluctuating shape at a finite rotational P\'eclet number, $\text{Pe} \simeq 90$, is, remarkably identical to the steady shape at $\text{Pe} = \infty$ (athermal simulations).  The sheet's shape is that of an `S'. This `S' shape has also been predicted for no-slip fibers,  along with other deformation modes \cite{wiens2015simulating}. However, the difference is that here the `S' shape is the time-averaged shape of a fluctuating object that does not perform full rotations, while in the case of fibers it is an instantaneous shape of a rotating object. Shape deformations in our simulation are due to the combination of thermal (Brownian) and flow-induced forces.

The S-shape is accurately predicted by an elastic beam model, where the load is extracted from the hydrodynamic load on a flat plate. This approximation is possible because the deformation is not large, a consequence of the fact that a slip sheet, unlike a no-slip sheet, never reaches the compressional axis of the flow \cite{gravelle2021violations,kamal2021alignment}.

We are also able to calculate the probability distribution function of the orientation of the midpoint of the sheet. The measured time-averaged orientation angle from MD corresponds approximately to the instantaneous orientation at infinite P\'eclet numbers of a flat slip plate. Importantly, the instantaneous shape fluctuations do not bring the particle outside the region of stable orientation. 

The results of this paper are comforting because they suggest that in the study of the orientational and deformation statistics of graphene, one can rely on the statistics of the hydrodynamic load on flat rigid slip plates, at least for sheets that are not too long.  
The successful application of a theory for rigid particles results from the particle relatively close from its equilibrium shape and orientation, which is the case for the particles considered in the current article. For highly deformed particles, like long graphene particles of linear dimension significantly larger than $1 \mu$m, or for small $\text{Pe}$, these assumptions may have to be reconsidered, particularly if the deformations induce the particle to rotate and explore the full orientational space.  Indeed, the stable hydrodynamic well in which the sheet orientation is trapped depends on the parameter $|k_\text{e}|$, which itself depends on the particle aspect ratio \cite{kamal2021effect}. Due to computational limitations, the behavior of long sheet-like particles can not be explored using all-atom MD. The use of coarse-grained atomic simulation models may help explore this regime in the future \cite{ye2019smoothed}. If the particle's orientation becomes unstable, we expect buckling or even chaotic tumbling orbits, similar to those of no-slip sheets \cite{dutta2017dynamics,silmore2021buckling,verhille2022deformability,funkenbusch2024dynamics} or fibers  \cite{liu2018morphological, du2019dynamics}.

Our calculations were performed for quasi-2D particles with their normals in the flow-gradient plane. An analysis of deformation in full 3D would be interesting, as it could give insights into torsional modes of deformation. The analysis of the trajectories of fully 3D ellipsoids is the subject of ongoing work. The current study quantifies the impact of the fluctuating shape of the particle on the orientation and average shape of the particle, two quantities that rheological models for dilute solutions of anisotropic flexible particles are sensitive to \cite{liu2018morphological, du2019dynamics}.  

\backmatter

\bmhead{Data availability statement}

The LAMMPS inputs for the MD simulations are available in the GitHub repository; \href{https://github.com/simongravelle/publication-data}{https://github.com/simongravelle/publication-data} \cite{simon_gravelle_2024_13341067}. The BI data is available upon request.

\bmhead{Acknowledgements}

S.G. acknowledges funding from the European Union's Horizon 2020 research and innovation programme under the Marie Skłodowska-Curie grant agreement N$^\circ\;101065060$, project NanoSep. C.K. acknowledges funding from the Royal Society Dorothy Hodgkin fellowship, grant number G119482. L.B. gratefully acknowledge funding from the European Research Council (ERC) under the European Union’s Horizon 2020 research and innovation program (grant agreement N$^\circ\;715475$, project FlexNanoFlow).

\bibliographystyle{sn-basic}
\bibliography{references}

\end{document}